\documentclass{article}

\usepackage{amssymb}
\usepackage{color}

\begin{document}

\title{\bf Quantum Black Hole in the Generalized Uncertainty Principle Framework}
\author{A. Bina$^1$\thanks{email: amir.bina82@gmail.com(a-bina@arshad.araku.ac.ir)},\,
S. Jalalzadeh $^{2,3}$\thanks{email: s-jalalzadeh@sbu.ac.ir},\,
\\A. Moslehi$^1$
\\$^1${\small Department of Physics, Faculty of Science, Arak University, Arak 879, Iran}
\\$^2${\small Department of Physics, Shahid Beheshti University G.C., Evin, Tehran 19839, Iran}
\\$^3${\small Research Institute for Astronomy and Astrophysics of Maragha
(RIAAM)- Maragha, Iran,P. O. Box: 55134-441}
}
\date{\today}
\maketitle
\begin{abstract}
In this paper we study the effects of the Generalized Uncertainty Principle (GUP) on canonical quantum gravity of black holes. Through the use of modified partition function that involves the effects of the GUP, we obtain the thermodynamical properties of the Schwarzschild black hole. We also  calculate the Hawking temperature and entropy for the modification of the Schwarzschild black hole in the presence of the GUP.

\end{abstract}
\vspace{1mm}\noindent
\section{Introduction}
\indent

The discovery of  temperature and  entropy of black holes is one of the most important achievements in  gravitational physics. Since  entropy is a statistical concern, it has been in a great interest \cite{Bek1}-\cite{Hawk2}. In the early seventies Bekenstein proposed the quantization of  black holes \cite{Bek1}. He showed that the gravitational surface of a black hole is proportional to its temperature and so the event horizon area is proportional to the entropy. He concluded that the event horizon of non-extremal black holes behaves like an adiabatic invariant, thus the event horizon should have discrete spectrum in the general relativity framework \cite{Hawk1}.

Similarities between these rules and  thermodynamics was first investigated by Hawking \cite{Hawk2}. Later on he discovered the evaporation of  black holes by a series of semi-classical calculations. It meant that the black holes like black bodies emit thermal radiation proportional to their gravitational surface. In addition the entropy is equivalent to one-forth of the event horizon area. Hawking's calculations introduced the relation between  classical mechanics of a black hole and its thermodynamics. These results led to deeper correspondence between classical gravity, quantum mechanics and statistical mechanics.

During these years the entropy of  black holes has been studied  by different approaches, for example by string theory, loop quantum gravity and canonical gravity \cite{Ashtekar}. Although theses methods,  involves curved geometry, we can perform  Hawking radiation calculations in a flat space. Also, up to now these calculations  have been done for extremal and near extremal
black holes \cite{Stro}.

The existence of a minimal length is one of the most interesting predictions of the theories related to  quantum gravity \cite{{Gross},{Kato}}. From the perturbative string theory this length is due to the fact that the strings cannot influence through distances smaller than their size. One interesting property of the existence of the minimal length is the modification of the standard commutation relation between position and momentum in usual quantum mechanics \cite{Kempf} which is called the generalized uncertainty principle (GUP). Noncommutativity between space-time coordinates was first studied by Snyder \cite{Snyder} and
\cite{Connes}-\cite{Dop}.

The noncommutativity theory is of great interest because of its interesting predictions in  particle physics, for example the mixture of IR/UV and non-locality \cite{Minwalla}, Lorenz violation \cite{Carroll}, canonical noncommutative with deformed
rather than broken symmetries {\cite{Fiore,G3}},  Lie-algebraic noncommutativity
\cite{{Marija},{G4}} and modern physics at small scales {\cite{Carroll,Szabo}}. Also in the past much attention years has been paid  to these fields (for applications of the GUP and non-commutativity in minisuperspace dynamics see \cite{frwnc}-\cite{BaMo2}).

While studying the thermodynamics of a black hole, the analysis of its temperature leads to many questions. When the initial state of a black hole is a pure quantum one and evolves to a mixed final state \cite{Hawk3}, first the black hole attracts  all the information behind its event horizon and then disappears through  thermal distribution. This causes the violation of the unitarity principle and  is associated with the information loss assumption. The uncertainty principle is one of the ways to escape this information loss \cite{page},\cite{Aharonov}. The Schwarzchild radius of a black hole with the Plank mass is of the order of the Plank length. Since this length is the wavelength of a particle with the Plank mass, if the mass of the black hole becomes lower than this mass, then we have a mass inside the volume smaller than that  allowed by the uncertainty principle. Zeldovich proposed that black holes with masses smaller than the
Plank mass are related to stable elementary particles \cite{Zel'dovich}.

There are many questions about minimal length during the study of black holes \cite{{medved},{G1},{G2},{Zhao}}. By investigation of the space-time for string scattering with an increasing higher orders in  perturbation theory, the size of the string reduces relative to the Schwarzchild radius of the collision region, thus the production of black holes becomes impossible in such a way but the length approach will make it possible but complicated \cite{Hossenfelder}. Recently string and loop quantum gravity theories  have succeeded to account for the entropy-event horizon area \cite{medved}.

In this paper in section 2, we  review a model to solve the Wheeler-Dewitt equation of the Schwarzschild black hole and  derive  its thermodynamics. In section 3, we examine Hilbert space representation in the generalized uncertainty principle framework. In section 4, we  study the Schwarzschild black hole with the generalized uncertainty relation and finally derive the thermodynamical properties of the quantum black hole.
\section{The Model}
\indent

The Wheeler-Dewitt equation for a Schwarzschild black hole  where the Hamiltonian involves only coordinates and momenta ($a, p_{a}$) i.e. $\mathcal{H}=\frac{p^{2}_a}{2a}+\frac{1}{2}a$ , can be obtained as \cite{Repo,obr sab T}
\begin{eqnarray}\label{2,1}
\frac{\hbar ^{2}G^{2}}{c^{6}}a^{-s-1}\frac{d}{da}\left( a^{s}\frac{d}{da}%
\psi (a)\right) =(a-\frac{2GM}{c^{2}})\psi (a),
\end{eqnarray}
where $a$ and $P_{a}^{2}= -\frac{\hbar ^{2}G^{2}}{c^{6}}a^{-s}\frac{d}{da}%
\left( a^{s}\frac{d}{da}\psi (a)\right) $ are phase coordinates deduced from the phase space coordinates $m$ and $P_{m}$, by means of an appropriate canonical transformation and also $m(t)=M(t,r)$ and
$P_{m}(t)=\int_{-\infty }^{\infty }drP_{M}(t,r)$. The variable $m$ can be defined as  mass $M$ of the hole when Einstein's equations are satisfied \cite{Repo} and $s$ is a factor ordering parameter. In particular, if we choose $s=2$ and identifying $R_{\it {s}}=\frac{2GM}{c^{2}}$, we have
\begin{eqnarray}\label{qe} \frac{\hbar ^{2}G^{2}}{c^{6}}\frac{1}{a}\left(
\frac{d^{2}}{da^{2}}+\frac{2}{a}\frac{d}{da}\right) \psi
(a)=(a-R_{\it {s}})\psi (a)~.
\end{eqnarray}
Let us consider the following transformations
\begin{eqnarray}\label{2,3}
\left\{
\begin{array}{lll}
\psi (a)=\frac{1}{a}\,\,U(a) \\
\\
\xi=a-R_{\it {s}},
\end{array}
\right.
\end{eqnarray}
where the variable $\xi$ indicates the gravitational degrees of freedom of the Schwarzschild black hole, and   define the appropriate constants and consider the fact that the energy of
excitations associated with  variable~ ${a}$ ~is not positive  \cite{obr sab T}. The
physical reason is simply that the total energy of the black hole is included and the ADM energy is equal to zero. Then, the quantum equation (\ref{qe}) turns into
\begin{eqnarray}\label{sbh}
\left( -\frac{\ell_{p}^{2}\,E_{p}}{2}\,\frac{d^{2}}{d{\xi}^{2}}+\frac{1}{2}\frac{E_{p}}{\ell_{p}^{2}}\,{\xi}^{2}\right) U(\xi)=\frac{R_{\it {s}}}{4\ell_{p}}\,E_{\it {s}}\,U(\xi),
\end{eqnarray}
where $E_{\it {s}}= Mc^{2}$ is the black hole ADM energy and $\ell _{p }=\sqrt{%
\frac{G\hbar }{c^{3}}}$ and $E_{p }=\sqrt{\frac{c^{5}\hbar}{G}}$. It  can
easily be shown that Eq.(\ref{sbh}) agrees with the Beckenstein's proposal \cite{Bek1} and represents a quantum linear oscillator  with energy levels of
\begin{eqnarray}\label{2,5}
 \frac{R_{\it {s}}(n)}{4\ell _{p}}\,E_{\it {s}}(n)=
 \left( n+\frac{1}{2}\right) E_{p}.
\end{eqnarray}
Also according to above equation one can get the mass of the black hole as
\begin{eqnarray}\label{m}
M^2(n)=\frac{2 \hbar c}{G} \left( n+\frac{1}{2}\right).
\end{eqnarray}
%

\subsection{Black Hole Entropy}
\indent

Details of the thermodynamical properties of a black hole can be obtained from its partition function \cite{obr sab T}-\cite{Kastrup}. For a quantum mechanical system the Feynman's path integral approach is a useful method
for determination of the free energy and partition function. In this way, to include the quantum effects \cite{{Feynman},{Kleinert}} a corrected potential in considered. In the case of the black hole with respect to Eq.(\ref{sbh}) the quantum equation is similar to the equation of a one-dimensional harmonic oscillator with  frequency of $\hbar \omega =\sqrt{\frac{3}{2\pi }}E_{p}$, and the corrected potential

\begin{eqnarray}\label{cv}
V(\xi)=\frac{3E_{p}}{4\pi \ell_{p}^{2}}\left({\xi}^{2}+\frac{\beta \ell_{p}^2E_{p}}{12}\right),
\end{eqnarray}
which leads to the following partition function \cite{{obr sab T},{obr sab}}
\begin{eqnarray}\label{2,1,2}
Z_{Q}=\sqrt{\frac{2\pi}{3}}~\frac{\exp (-\frac{\beta^{2}
E_{p}^{2}}{16\pi})}{\beta E_{p}}~.
\end{eqnarray}
From the fact that the internal energy of a black hole is equal to its gravitational energy, i.e.
\begin{eqnarray}\label{2,1,3}
\overline{E}=-\frac{\partial \ln (Z_{Q})}{\partial \beta }=Mc^{2},
\end{eqnarray}
we get
\begin{eqnarray}\label{2,1,4}
\frac{E_{p}^{2}}{8\pi }\beta ^{2}-Mc^{2}\beta +1=0~.
\end{eqnarray}
The positive solution for this equation when $E_{p}\ll Mc^{2}$ is
\begin{eqnarray}\label{2,1,5}
\beta =\beta _{H}\left[ 1-\frac{1}{\beta _{H}Mc^{2}}\right]~,
\end{eqnarray}
where $\beta _{H}=\frac{8\pi Mc^{2}}{E_{p }^{2}}=\frac{1}{kT_{H}}$,
is the Hawking's temperature. The entropy of the black hole using the partition
function and internal energy is defined as
\begin{eqnarray}\label{2,1,6}
\frac{\mathit{S}}{k}=\ln{Z_Q}+\beta\bar{E} ~.
\end{eqnarray}
Putting the Hawking's temperature in this equation we obtain
\begin{eqnarray}\label{2,1,7}
\frac{S}{k}=\frac{A_{s}}{4\ell_{p}^{2}}\left[ 1-\frac{1}{8\pi }\frac{
E_{p}^{2}}{\left( Mc^{2}\right) ^{2}}\right] ^{2}-\frac{1}{2}\ln \left(
\frac{A_{s}}{4\ell_{p}^{2}}\left[ 1-\frac{1}{8\pi }\frac{E_{p}^{2}}{
\left( Mc^{2}\right) ^{2}}\right] ^{2}\right) -\frac{1}{2}\ln (24)+1~,
\end{eqnarray}
where $A_{s}=4\pi R_{s}^{2}$ is the area horizon.

In terms of the Bekenstein-Howking relation ~$S_{BH}/{k}=A_{s}/{4\ell_{p}^{2}}$ ~ and ignoring terms of higher order, one can find the logarithmic correction to the entropy as is acquired  using different procedures in \cite{{Kastrup},{Gour}}
\begin{eqnarray}\label{2,1,8}
\frac{S}{k}=\frac{S_{BH}}{k}-\frac{1}{2}\ln \left( \frac{S_{BH}}{k}\right)
+\mathrm{O}\left({\mathit{S}^{-1}_{BH}}\right).
\end{eqnarray}
This result has the interesting feature that the coefficient of the first correction, the logarithmic one, agrees with the one obtained in loop quantum gravity \cite{loop Q BH}, as well as in  string theory \cite{Mukherji Pal}. The form of this correction  was already obtained  from other papers by other considerations \cite{{Gour},{Hod Vaz},Vladan}.

\section{Hilbert Space Representation in the GUP Framework}
\indent

In this section, we briefly study the modified Heisenberg algebra. In one dimension, deformation of the Heisenberg algebra generated by $\mathbf{X}$ , $\mathbf{P}$  is given by
\begin{eqnarray}\label{GUPJ}
[\mathbf{X},\mathbf{P}]=i\hbar (1+\sigma \mathbf{P}^2),\qquad \qquad\sigma > 0
\end{eqnarray}
where $\sigma$ is the deformation parameter and this commutation
relation  can be seen in perturbative string theory \cite{Gross}. The above equation by Kemf, Mangano and collaborators leads to the following relation
\cite{{kempf1},{kempf2}}
\begin{eqnarray}\label{GUPR}
\Delta x \Delta p \ge \frac{\hbar}{2} (1+\sigma (\Delta p)^2
+ \sigma \langle {\mathbf{P}} \rangle^2),
\end{eqnarray}
so that the canonical Heisenberg algebra is satisfied in the limit $\sigma \rightarrow 0$. By paying attention to the above equation the uncertainty in momentum  will be
\begin{eqnarray}\label{3,3}
\Delta p = \frac{\Delta x}{\hbar\sigma}  \pm
\sqrt{\left(\frac{\Delta x}{\hbar \sigma}\right)^2 -\frac{1}{\sigma}
- \langle{\mathbf{P}}\rangle^2}.
\end{eqnarray}
So the minimal uncertainty in position is $\langle {\mathbf{P}} \rangle$ dependent, that is
\begin{eqnarray}\label{Xmin}
\Delta x_{min}(\langle{\mathbf{P}}\rangle)
= \hbar \sqrt{\sigma} \sqrt{1+\sigma \langle{\mathbf{P}}\rangle^2}.
\end{eqnarray}
It is clear that the smallest uncertainty in position occurs when $\langle {\mathbf{P}}\rangle = 0$ and is equal to
$\Delta x_{min} = \hbar\sqrt{\sigma}$. This relation says that it is impossible to consider any physical state as the eigenstate of the position \cite{{kempf1},{kempf2}}, therefor in the presence of $\Delta x_{min}$, the definition of a state $\vert\psi_n\rangle\in \mathcal{D}$ ($\mathcal{D}\subset\mathcal{H}$ from a Hilbert space) such that
\begin{eqnarray}\label{3,5}
\lim_{n\rightarrow\infty}\left(\Delta x_{min}\right)_{\vert\psi_n\rangle}=
\lim_{n\rightarrow\infty}\langle\psi\vert({\bf \mathbf{X}}-\langle\psi\vert{\bf \mathbf{X}}\vert\psi\rangle)^2\vert\psi\rangle=0,
\end{eqnarray}
is impossible. It means that the eigenstates of the position are no longer physical states and should be assumed as formal states. Consequently in the GUP approach we cannot work in the space configuration and have to use ``quasi-position" states.

Eq.(\ref{GUPR}) involves both the low  and high energy regions which are related to quantum mechanics and quantum gravity limits respectively. These limits as a sample of applications of the GUP have been derived through string theory \cite{{Amati2},{Konishi}}. The quantum mechanical limit is given by
\begin{eqnarray}\label{q2}
 \sigma(\Delta p)^2+\sigma\langle p\rangle^2\ll\ 1 \mapsto (\Delta p)^2+\langle
 p\rangle^2\ll\frac{1}{\sigma}\;,
\end{eqnarray}
 Also the quantum gravity limit is of the form
\begin{eqnarray}\label{q3}
 \sigma(\Delta p)^2+\sigma\langle p\rangle^2\sim\ 1 \mapsto (\Delta p)^2+\langle
 p\rangle^2\sim\frac{1}{\sigma}.
\end{eqnarray}
%

\subsection{Representation in Momentum Space}
\indent

With respect to the lack of non vanishing minimal uncertainty in momentum the Heisenbergh algebra can be represented in momentum space wave function $\psi(p)=\langle p \vert \psi \rangle $. On a dense domain in the Hilbert space ${\mathbf{X}}$ and ${\mathbf{P}}$ play as operators such that \cite{{chang1},{dadic},{Bouaziz},{chang2}}
\begin{eqnarray}\label{ogup}
{\mathbf{P}}\psi(p) & = & p\;\psi(p),\nonumber  \\
{\mathbf{X}}\psi(p) & = & i\hbar \left[(1+\sigma p^{2})\frac{\partial}{\partial p}+\sigma p\right] \psi(p).
\end{eqnarray}
\smallskip\newline
As can be seen $\mathbf{X}$ and $\mathbf{P}$ are symmetric and this representation is easily seen to respect the commutation relation (\ref{GUPJ}). The scalar product of two arbitrary wave functions on the mentioned dense domain
 is given by

\begin{eqnarray}\label{3,1,2}
\langle \Phi |\Psi\rangle
=\int_{-\infty}^{+\infty} dp\, \Phi^*(p)\,\Psi(p)\;.
\end{eqnarray}
It should be noted that in the presence of  minimal uncertainty in position the momentum operator is still self-adjoint and also the functional analysis of the position operator changes.
Note that the definition of inner product and representation of $\mathbf{X}$ and $\mathbf{P}$ operators
in this paper are different from corresponding definition on Kemf, Mangano and collaborators
\cite{{kempf1},{kempf2}}. In fact they used the following representation for Hilbert Space of states
\begin{eqnarray}\label{3,1,3}
{\mathbf{P}}.\psi(p) & = & p\;\psi(p)\nonumber  \\
{\mathbf{X}}.\psi(p) & = & i\hbar (1+\sigma p^{2})\partial_{p}\psi(p)
\end{eqnarray}
And also

\begin{eqnarray}\label{3,1,4}
\langle \Psi |\Phi\rangle
=\int_{-\infty}^{+\infty} \frac{dp}{1+\sigma p^{2}}\, \Psi^*(p)\,\Phi(p)\;.
\end{eqnarray}
%

\subsection{Maximal Localization States}
\indent

As  mentioned before when there is uncertainty in position eigenstates the eigenstates of position are not physical states therefor in order to gain information about position we have to write the matrix elements of position operator in another basis e.g. momentum basis. This leads  to the study of states which are called ``maximal localization states''. As a result we consider the state $\vert \psi^{ml}_{{\chi}} \rangle $, maximally localized around position ${\chi}$ with the following properties
\begin{eqnarray}\label{3,2,1}
\langle \psi^{ml}_{{\chi}}\vert {\mathbf{X}} \vert \psi^{ml}_{{\chi}}\rangle = {\chi,}
\end{eqnarray}
and
\begin{eqnarray}\label{3,2,2}
(\Delta x)_{\vert \psi^{ml}_{{\chi}} \rangle } = \Delta x_{min}.
\end{eqnarray}
Paying attention to the smallest uncertainty in position and considering the standard deviation in uncertainty relation, for any state in the Heisenbergh algebra we have
\begin{eqnarray}\label{3,2,3}
\langle \psi \vert ({\mathbf{X}}-\langle {\mathbf{X}}\rangle)^2 - \left(
\frac{\vert\langle[{\mathbf{X}},{\mathbf{P}}]\rangle\vert}{2 (\Delta p)^2}\right)^2
({\mathbf{P}}-\langle{\mathbf{P}}\rangle)^2 \vert \psi \rangle  \ge 0,
\end{eqnarray}
which immediately implies
\begin{eqnarray}\label{3,2,4}
\Delta x \Delta p \ge \frac{\vert \langle[{\mathbf{X}},{\mathbf{P}}]\rangle\vert}{2}.
\end{eqnarray}
It is clear that if the state $\vert \psi \rangle $ obeys  $\Delta x \Delta p = \frac{\vert \langle[{\mathbf{X}},{\mathbf{P}}]\rangle\vert}{2}$ ~then it will  obey%
\begin{eqnarray}\label{3,2,5}
\left({\mathbf{X}} -\langle {\mathbf{X}} \rangle + \frac{\langle [{\mathbf{X}},{\mathbf{P}}]\rangle}{2(\Delta p)^2}
({\mathbf{P}}-\langle {\mathbf{P}} \rangle)\right)\vert \psi \rangle  = 0.
\end{eqnarray}
By solving the above equation  we determine the differential equation governing the maximal localization states
\begin{eqnarray}\label{3,2,6}
\left[ i\hbar (1+\sigma p^2)\partial_p - \chi
+i \hbar\sigma p + i \hbar \frac{1+\sigma (\Delta p)^2}{2(\Delta p)^2}
p\right] \psi(p) = 0.
\end{eqnarray}
Thus the maximal localization states are
given by%
\begin{eqnarray}\label{3,2,8}
\psi^{ml}_{\chi}(p) = \sqrt{\frac{2\sqrt{\sigma}}{\pi}}
(1+ \sigma p^2)^{-1}\exp({\frac{-i\chi}{\hbar \sqrt{\sigma}}\tan^{-1}(\sqrt{\sigma}p)}).
\end{eqnarray}
In the non-deformed case the plane waves in momentum space or Dirac $\delta$-function in position space are maximal localized states, but here  having deformation $\psi^{ml}_{\chi}(p)$ can be considered as a generalization of  plane waves.
\subsection{Transformation to Quasi-Position Wave Functions}
\indent

In order to investigate the probability of a state being in a maximally localized state around position ${\chi}$,
we consider the scalar product of  arbitrary states $\vert \phi \rangle$ on the states $\vert \psi^{ml}_{{\chi}} \rangle$  so that
$\phi({\chi}) = \langle \psi^{ml}_{{\chi}} \vert \phi \rangle$ is called the quasi-position wave function. Any wave function in the momentum representation can be transformed into its quasi-position counterpart by the following generalized Fourier transformation (this result is similar to Ref.\cite{{BaMo1},{kempf1}})
\begin{eqnarray}\label{gft}
\phi({\chi}) = \sqrt{\frac{2\sqrt{\sigma}}{\pi}} \int_{-\infty}^{+\infty}
\frac{dp}{1+\sigma p^2}\,\exp({\frac{i\chi}{\hbar \sqrt{\sigma}}\tan^{-1}(\sqrt{\sigma}p)})\, \phi(p).
\end{eqnarray}
Note that in the limit $\sigma \rightarrow 0$ the usual wave function $\phi({\chi}) = \langle {\chi} \vert \phi \rangle$ is determined and the above equation reduces to a plane wave in the momentum space.
\section{Quantum Black Hole with the Generalized Uncertainty Relation}
\indent

The commutation relation (\ref{GUPJ}) leads us to the generalized uncertainty relation (GUR) (\ref{GUPR}) \cite{chang1}-\cite{chang2}. According to Eq.(\ref{ogup}) the position and momentum operators are then represented in momentum space by
\begin{eqnarray}\label{4,1}
\left\{
\begin{array}{lll}
\mathbf{X}=i\hbar \left[(1+\sigma p^{2})\partial p+\sigma p\right], \\
\\
\mathbf{P}=p.
\end{array}
\right.
\end{eqnarray}
The Wheeler-DeWitt equation for the Schwarzschild black hole (\ref{sbh}) with the GUR then becomes
\begin{eqnarray}\label{4,2}
\Biggl\{-\frac{\hbar^{2}}{\ell_{p}^{2}}
\Biggl[\left((1+\sigma p^2)\partial p\right)^2
+2\sigma p\left((1+\sigma p^2)\partial p\right)
+2\sigma^2 p^2+\sigma\Biggr]+\frac{\ell_{p}^{2}}{\hbar^{2}}\,p^2
\Biggr\}\psi(p)=\frac{R_{\it{s}}}{2\ell_{p}{E_{p}}}\,E_{\it{s}}\,\psi(p)\;,
\end{eqnarray}
where $p$ is  canonical momenta conjugate to $\xi$. The exact solution
of  Eq.(\ref{4,2}) is given in \cite{{chang1},{dadic}}
\begin{eqnarray}\label{4,3}
\frac{R_{\it{s}}(n)}{4\ell_{p}}\,E_{\it{s}}(n) ={E_{p}}
\left[\left(n+\frac{1}{2}\right)\sqrt{1+\left(\frac{\sigma\hbar^{2}}
{2\ell_{p}^{2}}\right)^{2}}+\left(n^2+n+\frac{1}{2}\right)
\frac{\sigma\hbar^{2}}{2\ell_{p}^{2}}\right]\;.
\end{eqnarray}
and also for the mass of black hole we have
\begin{eqnarray}\label{gupm}
M^2(n)=\frac{2 \hbar c}{G} \left[\left(n+\frac{1}{2}\right)\sqrt{1+\left(\frac{\sigma\hbar^{2}}
{2\ell_{p}^{2}}\right)^{2}}+\left(n^2+n+\frac{1}{2}\right)
\frac{\sigma\hbar^{2}}{2\ell_{p}^{2}}\right]\;.
\end{eqnarray}
Now, we define
\begin{eqnarray}\label{4,4}
c  \;=\; \frac{1}{\sqrt{1+\sigma p^2}}\;,~~~~~~
s  \;=\; \frac{\sqrt{\sigma}p}{\sqrt{1+\sigma p^2}}\;,~~~~~~
2\nu \;=\; 1+\sqrt{1+\frac{4}{\sigma^2}},\;
\end{eqnarray}
\\
\\
the normalized energy eigenfunctions are
\begin{eqnarray}\label{sai}
\psi_n(p) =
2^{\nu}\Gamma(\nu)\sqrt{\frac{n!\,(n+\nu)\,\sqrt{\sigma}}
{2\pi\,\Gamma(n+2\nu)}}\;c^{\nu+1}\,C_n^{\nu}(s)\;,
\end{eqnarray}
where $n=0,1,2,\ldots$~and~ $C^{\nu}_n(s) $ is the \textit{Gegenbauer} polynomial
\\
\begin{eqnarray}\label{4,6}
C^{\nu}_n(s)=\frac{(-1)^n}{2^nn!}\frac{\Gamma(2\nu+n)\Gamma(\frac{2\nu+1}{2})}
{\Gamma(2\nu)\Gamma(\frac{2\nu+1}{2}+n)}(1-s^2)^{1/2-\nu}\frac{d^n}{ds^n}
(1-s^2)^{\nu+n-1/2}.
\end{eqnarray}
Also one can write (\ref{sai}) as \cite{abr}
\begin{eqnarray}\label{4,7}
\psi_n(p) ={2}^{\nu}\Gamma  \left( \nu \right)
\sqrt {{\frac {n!\, \left( n+1 \right) \sqrt {\sigma}}{2\pi \,\Gamma
 \left( n+2\,\nu \right) }}}~\frac{{c}^{\nu+1}\Gamma \left( n+2\,\nu\right)
{F(-n,n+2\,\nu;\,1/2+\nu;\,-1/2\,s+1/2)}}{\Gamma\left(n+1\right)\Gamma
\left( 2\,\nu\right)}.
\end{eqnarray}
Note that  using  Eq.(\ref{gft}), one can acquire $\psi_n({\chi})$
also according to the equations (\ref{q2}), (\ref{q3}) and  (\ref{gupm}) we conclude that in quantum gravity regime the mass of a black hole is proportional with the quantum number $n$ in contrast, according to Eq.(\ref{m}) with ordinary scales of energy in standard quantization of black holes, that mass is proportional to  $\sqrt{n}$ that it agrees with Beckenstein's proposal \cite{Bek1}.

\subsection{Black Hole Entropy with the GUR}
\indent

When there is not any deformation in  the system, the coordinates and momenta variables $x_{i}$ and $p_{j}$ are canonically conjugate i.e. $\{x_i,p_j\}=\delta_{ij}$~,~$\{x_i,x_j\}=\{p_i,p_j\}=0$, thus the thermodynamics of  the system is calculated by using the following partition function
\begin{eqnarray}\label{4,1,1}
Z=\frac{1}{2\pi\hbar}\,\int e^{-\beta\,H(x,p)}dxdp.
\end{eqnarray}
In the more general deformed case with the following generalized commutation relations
\begin{eqnarray}\label{4,1,2}
\left[X_i, P_j\right] = i\hbar ~f_{ij}(X,P),\nonumber \\
\left[P_i, P_j\right] = i\hbar ~h_{ij}(X,P),\nonumber \\
\left[X_i, X_j\right] = i\hbar ~g_{ij}(X,P).
\end{eqnarray}
where operators $X_{i}$ and $P_{j}$ are new coordinates and momentum variables
respectively and $f_{ij}$, $g_{ij}$ and $h_{ij}$ are the deformation functions that obey properties like bilineary, Libniz rules and Jacobi identity.  In the classical limit $\hbar\rightarrow0$ the above relations reduce to the deformed Poisson brackets
\begin{eqnarray}\label{4,1,3}
\{X_i,P_j\}=f_{ij}(X,P), \quad~ \{P_i,P_j\}= h_{ij}(X,P), \quad~
\{X_i,X_j\}=g_{ij}(X,P)\,.
\end{eqnarray}
These relations are anti-symmetric, bilinear and obey the Libniz rules and Jacobi identity \cite{chang2,Tkachuk}. Then  the partition function for deformed case can be interpreted in terms of $X$ and $P$ \cite{Fityo}
as%
\begin{eqnarray}\label{4,1,5}
Z_{{\it deformed}}=\frac{1}{2\pi\hbar}\,\int e^{-\beta\,H(X,P)}\frac{dXdP}{J}~.
\end{eqnarray}
According to  Eq.(\ref{GUPJ}), $g(X,P)=h(X,P)=0$ and $\{X,P\}=f(X,P)=1+\sigma P^{2}$, therefore the partition function of  the quantum black hole becomes
\begin{eqnarray}\label{4,1,6}
Z_{{\it GUP}}=\frac{1}{2\pi\hbar}\,\int dX \exp\left[-\beta\,{V(\mathbf{X})}\right] \int dP
\frac{ \exp\left(-\frac{\beta \,c\,\ell_{p}}{{2\,\hbar}}\,{P^2}\right)} {1+\sigma P^2}~.
\end{eqnarray}
Now by inserting the modified potential (\ref{cv}) , the corrected partition function will be
\begin{eqnarray}\label{4,1,7}
Z^{\it{GUP}}_{\it {Q}}=\frac{\ell_{p}}{\hbar}~\sqrt {\frac {\pi}{3\,\beta E_{p}\,\sigma}}~
\exp\left[{-\left({\frac {\beta^{2}{E_{p}}^{2}}{16\pi }}
+{\frac {c\,\ell_{p}\,\beta}{2\,\hbar\,\sigma}}\right)}\right]~{\Gamma} \left(\frac{1}{2}\,,\,\frac {c\,\ell_{p}\,\beta}{2\,\hbar\,\sigma}\right).
\end{eqnarray}
For the case of  $\frac {c\ell_{p}\beta}{2\hbar\sigma}\gg1$  the above equation leads to
\begin{eqnarray}\label{4,1,8}
Z^{\it{GUP}}_{\it {Q}}=\sqrt{\frac{2\pi}{3}}~\frac{\exp\left(-{\frac {\beta^{2}{E_{p}}^{2}}{16\pi}}-\frac{\hbar\,\sigma}{c\,\ell_{p}\,\beta}\right)}{\beta\,{E_{p}}}~
~.
\end{eqnarray}
Now similar to the non-deformed case we put, $\overline{E}=-\frac{\partial \ln (Z^{\it{GUP}}_{\it {Q}})}{\partial \beta }=\frac{{E_{p}}^2}{8\pi}\beta+\frac{1}{\beta}-\frac{\hbar}{c\,\ell_{p}\,\beta^{2}}\,\sigma=Mc^{2}$. Hence,  the temperature of  the quantum black hole in the GUP framework in term of the Hawking temperature
becomes%
\begin{eqnarray}\label{4,1,9}
\beta=\beta _{H}\left[ 1-\frac{1}{\beta _{H}Mc^{2}}+\frac{M E_{p}}
{\left(\beta _{H}Mc^{2}-1\right)\left(\beta _{H}Mc^{2}-2\right)}\,\,\sigma \right]~.
\end{eqnarray}
The entropy is accounted for as before and by using  the obtained temperature, it is given as follows
\begin{eqnarray}\label{4,1,10}
\frac{{S_{\it {GUP}}}}{k}&=&\frac{A_{s}}{4\ell_{p}^{2}}\left[1-\frac{1}
{\beta_{H}Mc^{2}}\right]^{2} + \frac{A_{s}}{4\ell_{p}^{2}} \left[1-\frac{1}{\beta_{H}Mc^{2}}\right]\frac{M\,E_{p}}{
\left(\beta _{H}Mc^{2}-2\right)\left(\beta_{H}Mc^{2}-1\right)}\,\sigma
\nonumber\\
&&-\frac{1}{2}\ln\left(\frac{A_{s}}{4\ell_{p}^{2}}\left[1-\frac{1}
{\beta_{H}Mc^{2}}\right]^{2} + \frac{A_{s}}{4\ell_{p}^{2}} \left[1-\frac{1}{\beta_{H}Mc^{2}}\right]\frac{M\,E_{p}}{
\left(\beta _{H}Mc^{2}-2\right)\left(\beta_{H}Mc^{2}-1\right)}\,\sigma\right)
\nonumber\\
&&-\frac{2\,E_{p}\,\sigma}{c^{2}}\left[\beta_{H}
\left(1-\frac{1}{\beta_{H}Mc^{2}}\right)\right]^{-1}-\frac{1}{2}\ln(24)+1~.
\end{eqnarray}
Finally the definition of  GUP Hawking-Bekenstein entropy
\begin{eqnarray}\label{4,1,11}
\frac{S^{\it {GUP}}_{\it{BH}}}{k}=\frac{S_{\it{BH}}}{k}
\left(1+\frac{E_{p}^{3}}{8\pi\,M^{2}{c}^{6}}\,\sigma\right),
\end{eqnarray}
leads
to%
\begin{eqnarray}\label{4,1,12}
\frac{{S^{\it {GUP}}}}{k}=\frac{S^{\it {GUP}}_{\it {BH}}}{k}-\frac{1}{2}\,\ln \left( \frac{S^{\it {GUP}}_{\it {BH}}}{k}\right)-2Mc^{2}\left(\frac{S^{\it {GUP}}_{\it {BH}}}{S_{BH}}-1\right)+\mathrm{O}\left({\mathit{S}^{\it {GUP}\,\,-1}_{\it {BH}}}\right)~.
\end{eqnarray}
As the log-type correction is similar to the existing results that are derived from other methods \cite{{medved},{G2},{Zhao}} . It is shown that this result has the same form as the non-deformed case, again the logarithmic correction to the entropy appears with a $-1/2$ factor and it is clear that we get the non-deformed entropy in the limit $\sigma\rightarrow0$. As we can see, these thermodynamical quantities are modified due to the presence of the deformation parameter $\sigma$. In particular the GUP lessens the value of the entropy, which can be understood from the fact that  GUP reduces the accessible physical states. Similar results can be reached by surveying noncommutativity  \cite{{obr sab},{Bastos}}.
\section{Conclusion}
\indent

In summary,~we first introduced a model for quantum black holes and showed that its Wheeler-Dewitt equation is similar to the equation of a one-dimensional harmonic oscillator. We then reviewed the thermodynamical properties of  quantum black holes for which the logarithmic correction of the entropy with a~$-1/2$~factor  appeared. Next we presented the Hilbert space in the generalized uncertainty principle framework and  obtained the relevant  generalized Fourier transformation which gives the quasi-position wave function from momentum space. In the next step we studied the quantum black hole in this framework and obtained its wave functions and energy eigenvalues and argued that in a quantum gravity regime the mass of a black hole is proportional to the integer $n$, in contrast to ordinary scales of energy in standard quantization of black hole where the mass is proportional to  $\sqrt{n}$. Finally we determined the thermodynamical properties in this scenario  by introducing
the relevant  Bekenstein-Hawking entropy in the GUP framework and concluded that again the logarithmic correction of the entropy appears with the factor  ~$-1/2$ ~and also the value of the entropy diminishes which can be comprehended from the fact that the GUP  reduces the available physical states .

\vskip 1truecm \centerline{\bf Acknowledgmentsan}  We would like to thank
H. R. Sepangi for a
careful reading of the manuscript and A.B. is grateful to M. Poorakbar for helpful discussions. This work has been supported by the Research Institute for Astronomy and Astrophysics of
Maragha, Iran. We also would like to thank the anonymous referee for valuable
comments to improve the paper.
\smallskip\newline

\end{document}